\documentclass[showpacs,twocolumn]{revtex4-1}
\usepackage[english]{babel}
\usepackage{graphicx}
\usepackage{amsmath, amsthm, amssymb}
\usepackage{dcolumn}
\usepackage{subfigure}
\usepackage{lipsum}
\usepackage{amssymb}
\usepackage{amsmath}

\begin{document}

\title{\bf Open-quantum-systems approach to complementarity in neutral-kaon interferometry}

\author{Gustavo de Souza $^a$}\email[]{\emph{Corresponding author: }gdesouza@iceb.ufop.br}
\author{J.G.G. de Oliveira Jr. $^b$}%\email[]{jgojunior@uesc.br}
\author{Adalberto D. Varizi $^c$}%\email[]{adalbertovarizi@gmail.com}
\author{Edson C. Nogueira $^c$}%\email[]{edsoncezar16@gmail.com}
\author{Marcos D. Sampaio $^c$}%\email[]{msampaio@fisica.ufmg.br}

\affiliation{$^a$ Universidade Federal de Ouro Preto -- Departamento de Matem\'{a}tica -- ICEB \\
	Campus Morro do Cruzeiro, s/n, 35400-000, Ouro Preto/MG -- Brazil}
\affiliation{$^b$ Universidade Estadual de Santa Cruz -- Departamento de Ci\^{e}ncias Exatas e Tecnol\'{o}gicas \\ 45662–900, Ilh\'{e}us/BA -- Brazil,}
\affiliation{$^c$ Universidade Federal de Minas Gerais -- Departamento de F\'{\i}sica -- ICEX \\ P.O. BOX 702,
	30.161-970, Belo Horizonte/MG -- Brazil}

\begin{abstract}
In bipartite quantum systems, entanglement correlations between the parties exerts direct influence in the phenomenon of wave-particle duality.~This effect has been quantitatively analyzed in the context of two qubits by M. Jakob and J. Bergou [Optics Communications 283(5) (2010) 827].~Employing a description of the $K$-meson propagation in free space where its weak decay states are included as a second party, we study here this effect in the kaon-antikaon oscillations.~We show that a new quantitative ``triality'' relation holds, similar to the one considered by Jakob and Bergou.~In our case, it relates the distinguishability between the decay products states corresponding to the distinct kaon propagation modes $K_S $, $K_L $, the amount of wave-like path interference between these states, and the amount of entanglement given by the reduced von Neumann entropy.~The inequality can account for the complementarity between strangeness oscillations and lifetime information previously considered in the literature, therefore allowing one to see how it is affected by entanglement correlations.~As we will discuss, it allows one to visualize clearly through the $K^{0}$--$\overline{K}\,^{0}$ oscillations the fundamental role of entanglement in quantum complementarity.
%Lorem ipsum dolor sit amet, consectetuer adipiscing elit. Ut purus elit, vestibulum ut, placerat ac, adipiscing vitae, felis. Curabitur dictum gravida mauris. Nam arcu libero, nonummy eget, consectetuer id, vulputate a, magna. Donec vehicula augue eu neque. Pellentesque habitant morbi tristique senectus et netus et malesuada fames ac turpis egestas. Mauris ut leo. Cras viverra metus rhoncus sem. Nulla et lectus vestibulum urna fringilla ultrices.
\end{abstract}

\maketitle
%Prepared for submission to, for example, European Physical Journal C: Particles and Fields (they accept Letters too).

\section{Introduction}
\label{sec:intro}

\par In the present work, we revisit the complementarity between strangeness oscillations and lifetime information in the neutral kaon system previously studied by A. Bramon, G. Garbarino and B. Hiesmayr (BGH) \cite{BGH_PRL,BGH_TwoPath,BGH_EPJC}. However, instead of using the Wigner--Weisskopf approach to the isolated, free-kaon propagation, we consider an open systems model in which the neutral kaon's weak decay states are included as a second party. The interaction between the two subsystems is given by a completely positive probability-preserving quantum dynamical map. Our model coincides with that proposed by Caban \emph{et al}. \cite{Caban_Phys_Review_A} (and also discussed in Bertlmann \emph{et al}. \cite{BGH_OpenSys}) upon partial trace of the decay products, but it has the new feature of allowing bipartite entanglement to be studied. We examine quantitatively the effects of these correlations on complementarity in the context of neutral kaon interferometry.

\par In this case, the quantitative duality relation of the Greenberger-Yasin type \cite{Greenberger_Yasin} considered in Refs. \cite{BGH_PRL,BGH_TwoPath,BGH_EPJC} must extend to a ``triality'' relation incorporating a quantitative entanglement measure. We show here that a new such quantitative complementarity relation holds:

\vspace{-0.2cm}
\begin{equation}\label{eq:fidelity_compl}
\mathcal{V}(\tau ) \leq \sqrt{1-\mathcal{D}^2 (\tau ) - \mathcal{S}^2 (\tau )}\quad \forall \tau\in I ,
\end{equation}

\noindent where $\tau $ denotes the proper time and $I$ is the time interval relevant for the analysis (see Sec. \ref{sec:model}). This inequality is similar to that proposed by M. Jakob and J. Bergou for bipartite systems \cite{Jakob_Bergou}. Here, $\mathcal{D}$ denotes the distinguishability between the decay products states corresponding to the distinct kaon propagation modes $K_S $, $K_L $. As we will see, $\mathcal{D}$ quantifies the increasing amount of lifetime information which becomes available (due to entanglement correlations) in the decay states subsystem. The associated visibility $\mathcal{V}$ quantifies the amount of wave-like path interference between these states, while $\mathcal{S}$ denotes the von Neumann entropy of the kaon state and measures bipartite entanglement.~We will demonstrate that the new quantitative complementarity relation (\ref{eq:fidelity_compl}) also accounts for the complementarity between strangeness oscillations and lifetime information considered by BGH. The results allow us to visualize and discuss in a clear way through the $K^{0}$--$\overline{K}\,^{0}$ oscillations the essential role played by entanglement in wave-particle duality.

%%%%%%%%%%%%%%%%%%%%%%%%%%%%%%%%%%%%%%%%%%%%%%%%%%%%%%%%%%%%%%%%%%%%%%%%%%%%%%%%%%%%%%%%%%%%%%%%%%%%%%%%%%%%%%%%%%%%%%%%%%%%%%%%%%%%%%%%%%%%%%
\vspace{-0.3cm}
\section{The Model}
\label{sec:model}

\par While there are several open quantum system models available in the literature offering completely-positive, probability preserving descriptions of the composite neutral kaon plus weak decay products system \cite{Caban_Phys_Review_A, BGH_OpenSys, Caban_Phys_Lett_A, Caban_Phys_Lett_A_2, Smolinski_1, Smolinski_2}, here we consider a model in which these two subsystems are treated as different parties. Therefore, we take the composite system state space as the tensor product $\mathcal{H}=\mathcal{H}_Q \otimes \mathcal{H}_P $ between the kaon (quanton) Hilbert space $\mathcal{H}_Q $ and the space of decay products states $\mathcal{H}_P $.

\par A short-lived kaon $K^{0}_S $ always decays into two pions, either $\pi ^{+} + \pi ^{-}$ or $\pi ^{0} + \pi ^{0}$. On the other hand, a long-lived kaon $K^{0}_L $  has several decay modes: it can decay into three neutral pions or $\pi ^{+} + \pi^{-} + \pi ^{0}$, but there are also the semileptonic decays into $\pi ^{\pm } + \mu ^{\mp }+ \nu _{\mu }$, $\pi ^{\pm } + e^{\mp } + \nu _{e }$, and the considerably rare $K^{0}_L $ decays into two pions. However, we will not consider here this last decay mode associated with charge-parity violation. In this case, the state of the decay products subsystem can be labeled by its pion content. Thus, we take $\mathcal{H}_P $ as the Hilbert space spanned by the (orthonormalized) vectors $\arrowvert 0_\pi \rangle $, $\arrowvert\pi\pi\rangle$, and $\arrowvert\widetilde{\pi\pi}\rangle $, which represent respectively states with no pions, two pions, and one or three pions.

\par The kaon state space is taken as the direct sum $\mathcal{H}_Q = H_{0 }\oplus H_{K^0 }$, where $H_{0 }$ is the Hilbert space spanned by the vector $\arrowvert 0_{K}\rangle $ representing the vacuum (absence of kaon) and $H_{K^0 }$ is the usual kaon Hilbert space spanned by the strangeness eigenstates $\arrowvert K^{0} \rangle $, $\arrowvert \overline{K}^{0}\rangle $. Under our assumption of charge-parity symmetry, the neutral kaon mass eigenstates $\arrowvert K_{S}^{0}\rangle $, $\arrowvert K_{L}^{0}\rangle $ corresponding to the short-lived and long-lived propagation modes are

\vspace{-0.3cm}
\begin{align}\label{eq:kaon_basis}
\arrowvert K_{S}^0 \rangle=\frac{1}{\sqrt{2}}(\arrowvert K^0 \rangle +\arrowvert \overline{K}^0 \rangle),\quad 
\arrowvert K_{L}^0 \rangle=\frac{1}{\sqrt{2}}(\arrowvert K^0 \rangle -\arrowvert \overline{K}^0 \rangle)\, ,
\end{align}

\noindent and we have $\langle K_{S}^0 \arrowvert K_{L}^0 \rangle = 0$. We assume that $\arrowvert 0_{K}\rangle $ is normalized and orthogonal to $\arrowvert K^{0}_S \rangle $, $\arrowvert K^{0}_L \rangle $.

\vspace{0.2cm}
\par So far for the kinematic aspects. Let us turn now to dynamics. The only physically meaningful initial configurations are those with a kaon and no pion -- that is, factorized initial conditions of the form $\arrowvert \Psi (0) \rangle = \left(\, \alpha \arrowvert K_{S}^{0}\rangle + \beta \arrowvert K_{L}^{0}\rangle\, \right)\arrowvert 0_\pi \rangle$. We assume that evolution takes place entirely in the subspace $\mathcal{W}<\mathcal{H} $ spanned by $\{\arrowvert K_S^0 \rangle\arrowvert 0_\pi \rangle , \arrowvert K_L^0 \rangle\arrowvert 0_\pi \rangle , \arrowvert 0_{K}\rangle\arrowvert\pi\pi\rangle,
\arrowvert 0_{K}\rangle\arrowvert\widetilde{\pi\pi}\rangle \}$ and according to the quantum map

%\vspace{-0.2cm}
\begin{widetext}
\begin{align}\label{eq:dyn_map}
\arrowvert 0_{K}\rangle \arrowvert 0_\pi \rangle &\longrightarrow \arrowvert 0_{K}\rangle \arrowvert 0_\pi \rangle \\ \nonumber 
\arrowvert K_S^0 \rangle\arrowvert 0_\pi \rangle &\longrightarrow e^{-\tfrac{1}{2}\Gamma _S \tau }e^{-im_S \tau }\arrowvert K_S^0 \rangle\arrowvert 0_\pi \rangle + \sqrt{1-e^{-\Gamma _S \tau }}\arrowvert 0_{K}\rangle\arrowvert\pi\pi\rangle \\ \nonumber
\arrowvert K_L^0 \rangle\arrowvert 0_\pi \rangle &\longrightarrow e^{-\tfrac{1}{2}\Gamma _L \tau }e^{-im_L \tau }\arrowvert K_L^0 \rangle\arrowvert 0_\pi \rangle + \sqrt{1-e^{-\Gamma _L \tau }}\arrowvert 0_{K}\rangle\arrowvert\widetilde{\pi\pi}\rangle\; ,
\end{align}
\end{widetext}

\noindent where $m_S $ and $\Gamma _S = \frac{1}{\tau _S }$ (resp. $m_L $ and $\Gamma _L = \frac{1}{\tau _L }$) are the $K^0_S $ (resp. $K^0_L $) mass and decay width \cite{footnote_1}, and where $p_S (\tau )\equiv 1-e^{-\Gamma _S \tau } $ (resp. $p_L (\tau )\equiv 1-e^{-\Gamma _L \tau } $) denotes the amplitude for the state $\arrowvert K_S^0 \rangle\arrowvert 0_\pi \rangle $ (resp. $\arrowvert K_L^0 \rangle\arrowvert 0_\pi \rangle $) mapping at time $\tau $ into $ \arrowvert 0_{K}\rangle\arrowvert\pi\pi\rangle $ (resp. $ \arrowvert 0_{K}\rangle\arrowvert\widetilde{\pi\pi}\rangle $). Moreover, we assume that all the interactions experienced by the kaon with other degrees of freedom have been included in the description above. In this case the composite system density operator $\rho(\tau )=\arrowvert \Psi (\tau ) \rangle \langle \Psi (\tau ) \arrowvert $ for the initial $\arrowvert \Psi (0) \rangle $ can be assumed to remain pure in the course of dynamics. From Eq. (\ref{eq:dyn_map}),

\vspace{-0.2cm}
\begin{widetext}
\begin{align}\label{eq:density}
\arrowvert \Psi (\tau ) \rangle &= \alpha e^{-\tfrac{1}{2}\Gamma _S \tau }e^{-im_S \tau }\arrowvert K_S^0 \rangle\otimes\arrowvert 0_\pi \rangle + \beta e^{-\tfrac{1}{2}\Gamma _L \tau }e^{-im_L \tau }\arrowvert K_L^0 \rangle\otimes\arrowvert 0_\pi \rangle + \alpha \sqrt{1-e^{-\Gamma _S \tau }}\arrowvert 0_{K}\rangle\otimes\arrowvert\pi\pi\rangle \\ \nonumber 
&+ \beta \sqrt{1-e^{-\Gamma _L \tau }}\arrowvert 0_{K}\rangle\otimes\arrowvert\widetilde{\pi\pi}\rangle \; .
\end{align}
\end{widetext}

\par In the sequence, we will focus our attention in kaons produced in strangeness eigenstates -- say, $\arrowvert K^{0} \rangle $ states generated by strong reactions such as $\pi^{-} p \rightarrow K^0 \Lambda $. So we take $\alpha = \frac{1}{\sqrt{2}} = \beta $. Taking this into $\rho (\tau )$, we see that the reduced kaon state is

%\vspace{-0.15cm}
\begin{align}\label{eq:reduced_kaon}
&\rho _Q (\tau ) = \mathrm{Tr}_{P}\left[ \rho (\tau ) \right] = \frac{1}{2}e^{-\Gamma _S \tau } \arrowvert K_S^0 \rangle \langle K_S^0 \arrowvert \\ \nonumber
&+ \frac{1}{2}e^{-\Gamma _L \tau } \arrowvert K_L^0 \rangle \langle K_L^0 \arrowvert + \left( 1 - \frac{e^{-\Gamma _S \tau }}{2}-\frac{e^{-\Gamma _L \tau }}{2}\right)\arrowvert 0_K \rangle \langle 0_K \arrowvert \\ \nonumber
&+ \left\{ \frac{1}{2}e^{-\frac{1}{2}\Gamma _S \tau }e^{-\frac{1}{2}\Gamma _L \tau }e^{i\Delta m \tau }\arrowvert K_S^0 \rangle \langle K_L^0 \arrowvert + \mathrm{H.C.}\right\}\; 
\end{align}

\noindent This coincides with the kaon state evolution considered in Ref. \cite{Caban_Phys_Review_A} by Caban \emph{et al}.~\cite{footnote_2}. In this work, the authors deduced the general form in (\ref{eq:reduced_kaon}) for the kaon's dynamics under the assumptions that the kaon state evolution must be (i) completely positive and probability preserving, and (ii) compatible with the Wigner--Weisskopf phenomenological prescription \cite{footnote_3}.~Therefore, these properties are also true for the reduced kaon state $\rho _Q (\tau ) $ in the present model (\ref{eq:density}).~It is straightforward to check that the composite system evolution given by $\rho (\tau )$ is also completely positive and probability preserving.

%%%%%%%%%%%%%%%%%%%%%%%%%%%%%%%%%%%%%%%%%%%%%%%%%%%%%%%%%%%%%%%%%%%%%%%%%%%%%%%%%%%%%%%%%%%%%%%%%%%%%%%%%%%%%%%%%%%%%%%%%%%%%%%%%%%%%%%%%%%%%%
\vspace{-0.2cm}
\section{Results}
\label{sec:model}

\par We apply our model now to a quantitative analysis of complementarity in the $K^{0}$--$\overline{K}\,^{0}$ system, including the duality between strangeness oscillations and lifetime information. But here we investigate the phenomenon in light of the new feature presented by the model's bipartite character: entanglement. Our goal is to examine its role on complementarity in the context of neutral kaon interferometry.

\par We can restrict our analysis to focus only on the proper time interval $0\leq \tau \leq \tau _0 $, where $\tau _0 = 4.79 \tau _S $. The reason is that it can be verified in experiment \cite{BGH_PRL} that neutral kaons decaying after $\tau _0 $ can be regarded as $K_0^L $ kaons with negligible error probability. In other words: at $\tau = \tau _0 $ one can already consider to have complete width information on the kaon. Therefore, we assume in the sequence that $\tau $ ranges from $0 $ to $\tau _0 $.

\par For pure composite system states, the degree of mixedness of a reduced party state both qualifies and quantifies entanglement. Here we will use the von Neumann entropy $\mathcal{S}$ of the reduced pionic subsystem state. It can be readily evaluated from

\begin{align}\label{eq:reduced_pion}
&\rho _P (\tau ) = \mathrm{Tr}_{Q}\left[ \rho (\tau ) \right] = \frac{1-e^{-\Gamma _S \tau } }{2} \arrowvert\pi\pi\rangle\langle\pi\pi\arrowvert \\ \nonumber
&+\frac{1-e^{-\Gamma _L \tau } }{2} \arrowvert\widetilde{\pi\pi}\rangle\langle\widetilde{\pi\pi}\arrowvert +\left( \frac{e^{-\Gamma _S \tau }}{2}+\frac{e^{-\Gamma _L \tau }}{2} \right) \arrowvert 0_{\pi }\rangle\langle 0_{\pi }\arrowvert \\ \nonumber
&+\left\{ \frac{1}{2}\sqrt{1-e^{-\Gamma _S \tau }}\sqrt{1-e^{-\Gamma _L \tau }}\arrowvert \pi\pi \rangle\langle \widetilde{\pi\pi }\arrowvert + \mathrm{H.C.} \right\}\; ,
\end{align}
\vspace{0.2cm}

\noindent whose eigenvalues are $\{ 0, x, 1-x \}$ for $$x(\tau )\equiv \dfrac{e^{-\Gamma _S \tau }+e^{-\Gamma _L \tau }}{2}\; ,$$ \noindent $0\leq x \leq 1 $ $\forall \tau $. Direct numerical analysis reveals that $$\mathcal{S}=x\ln x + (1-x)\ln (1-x)$$ \noindent is monotone increasing in $[0, \tau _0 ]$ (see Fig.~1).

\par As entanglement correlations are dynamically generated, information about the kaon's $\{ \arrowvert K_{S}^{0}\rangle , \arrowvert K_{L}^{0}\rangle \}$ component leaks to the pionic subsystem. The natural quantifier of the amount of lifetime information which thus becomes available to be retrieved (through the pionic state) is the distinguishability

\begin{equation}\label{eq:disting}
\mathcal{D}(\tau ) = \frac{1}{2}\| \rho _{P}^{(S)}(\tau ) - \rho _{P}^{(L)}(\tau )   \| \, .
\end{equation}
\vspace{0.1cm}

\noindent It is given by the trace distance between the pionic subsystem states $\rho _{P}^{(S)}(\tau ) = 2 \langle K_{S}^0 \arrowvert \rho (\tau ) \arrowvert K_{S}^0 \rangle = e^{-\tau \Gamma _S }\arrowvert 0_{\pi }\rangle\langle 0_{\pi }\arrowvert $ and $\rho _{P}^{(L)}(\tau ) = 2\langle K_{L}^0 \arrowvert \rho (\tau ) \arrowvert K_{L}^0 \rangle = e^{-\tau \Gamma _L }\arrowvert 0_{\pi }\rangle\langle 0_{\pi }\arrowvert $ corresponding to the distinct kaon propagation modes $K_S $, $K_L $. Due to the generation of entanglement, we expect $\mathcal{D}(\tau )$ to also increase monotonically in $[0,\tau _0 ]$. In fact, we found that $\mathcal{D}(\tau )=\frac{1}{2}\left|e^{-\tau \Gamma _S }-e^{-\tau \Gamma _L }\right|$ is an increasing function of $\tau $ in this interval. The numerical results are summarized in Fig.~1.

%%%%%%%%%%%%%%%%% Fig. 1
\begin{figure}[htp]
	\begin{center}
		\includegraphics[scale=0.38]{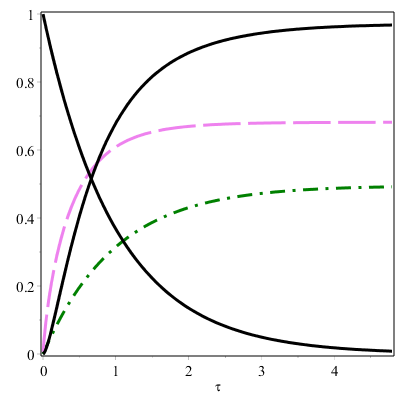}
		\caption{(color online) The solid black lines show $\mathcal {D}^2 + \mathcal {S}^2 $ increasing towards $1$ while $\mathcal{V}^2 $ decreases correspondingly, such that $\mathcal{V}^2 \leq 1 - \mathcal {D}^2 - \mathcal {S}^2 $. Proper time ranges from $\tau =0$ to $\tau = 4.79 \tau _S $. The green dotdashed line shows $\mathcal{D}$. The purple dashed line shows $\mathcal{S}$.}\label{fig1}
	\end{center}
\end{figure}
%%%%%%%%%%%%%%%%%

\par The quantity naturally complementary to $\mathcal{D}$ and playing the role of interferometric visibility here is the Uhlmann fidelity 

\begin{equation}\label{eq:vis_fidelity}
\mathcal{V}(\tau ) = \mathcal{F}\left( \rho _{P}^{(S)}(\tau ) , \rho _{P}^{(L)}(\tau ) \right)\, ,
\end{equation}

\noindent where $\mathcal{F}\left( \hat{\rho }, \hat{\tau }\right)=\mathrm{Tr}\left[ \displaystyle\sqrt{\sqrt{\hat{\rho }}\, \hat{\tau }\sqrt{\hat{\rho }}}\right]$. The fidelity is an ``overlap'' measure generalized to arbitrary mixed states $\hat{\rho }, \hat{\tau }$, therefore quantifying the visibility of quantum interferences between $\rho _{P}^{(S)}(\tau ), \rho _{P}^{(L)}(\tau ) $. Moreover, it is well-known to be related to the trace distance by the information-theoretic inequality

\begin{equation}\label{eq:inf_inequality}
\mathcal{F}(\hat{\rho }, \hat{\tau })\leq \displaystyle\sqrt{1-\mathcal{D}^2 (\hat{\rho }, \hat{\tau })}\; .
\end{equation}
%\vspace{0.2cm}

\noindent We have $\mathcal{V}(\tau )=e^{-\Gamma \tau }$, where $\Gamma \equiv \frac{1}{2}(\Gamma _S + \Gamma _L )$.

\par As was pointed out by Jakob and Bergou in Ref.~\cite{Jakob_Bergou}, complementarity in bipartite systems must relate the single-partite properties distinguishability and visibility to the amount of entanglement. Here we have

\vspace{-0.2cm}
\begin{equation*}
\mathcal{V}^2 +\mathcal{D}^2 = e^{-\tau (\Gamma _S + \Gamma _L )}+\frac{\left( e^{-\Gamma _S \tau }-e^{-\Gamma _L \tau } \right) ^2 }{4}=x (\tau ) ^2 \; ,
\end{equation*}

\noindent in such a way that

\vspace{-0.2cm}
\begin{align*}
\mathcal{V}^2 +\mathcal{D}^2 +\mathcal{S}^2 = x^2 + [x\ln (x) + (1-x)\ln (1-x)]^2 \leq 1\, .
\end{align*}

\noindent Indeed, numerical analysis of the quantities $\mathcal {S}$, $\mathcal {D}$ and $\mathcal {V}$ shows that the inequality

\begin{equation}\label{eq:bergou}
\mathcal {V}^2 + \mathcal {D}^2 + \mathcal {S}^2 \leq 1
\end{equation}

\noindent holds within the relevant proper time interval $[0,\tau _0 ]$ (see Fig. 2). As $\mathcal {D}^2 + \mathcal {S}^2 $ increases towards (nearly) $1$ in $[0, \tau _0 ]$, the quantity $\mathcal{V}$ must correspondingly decrease towards $0$, therefore enforcing the visibility of the wave-like, path interference phenomena to reduce.

\vspace{0.5cm}

%%%%%%%%%%%%%%%%% Fig. 2
%\begin{figure}[htp]
%	\begin{center}
%		\includegraphics[scale=0.35]{fig2a.png} \quad %\includegraphics[scale=0.35]{fig2b.png}
%        \caption{(color online) (a) The quantitative %complementarity relation (\ref{eq:bergou}). The solid violet %line shows $\mathcal {V}^2 + \mathcal {D}^2 + \mathcal {S}^2 $. %The upper bound $1$ is shown dashed, in black. (b) %Hi.}\label{fig2}
%	\end{center}
%\end{figure}
%\vspace{-0.5cm}
%%%%%%%%%%%%%%%%%
%%%%%%%%%%%%%%%%% Fig. 2
\begin{figure}[htp]
	\begin{center}
		\includegraphics[scale=0.38]{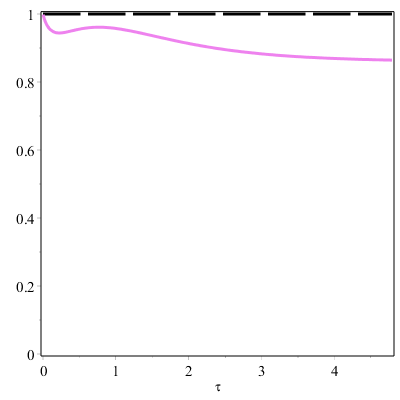}
		\caption{(color online) The quantitative complementarity relation (\ref{eq:bergou}) for $\tau \in [0,\tau _0 ]$. The solid violet line shows $\mathcal {V}^2 + \mathcal {D}^2 + \mathcal {S}^2 $. The upper bound $1$ is shown dashed, in black.}\label{fig2}
	\end{center}
\end{figure}
%%%%%%%%%%%%%%%%%%

\vspace{-1.0cm}

%%%%%%%%%%%%%%%%%%%%%%%%%%%%%%%%%%%%%%%%%
\subsection*{Strangeness Oscillations}
\label{subsec:strangeness}

\par To see that Eq.~(\ref{eq:bergou}) also accounts for the complementarity between strangeness oscillations and lifetime information in $[0,\tau _0 ]$, notice first that the visibility of $K^{0}\overline{K}^{0}$ oscillations must be defined here as the quantity $\mathcal{V}_0 (\tau )$ such that

\begin{equation}\label{eq:vis_strange}
2 \langle \overline{K}^{0} \arrowvert \mathrm{Tr}_{P} [\rho (\tau )] \arrowvert \overline{K}^{0} \rangle = F(\tau )\{ 1-\mathcal{V}_0 (\tau )\cos (\Delta m \tau ) \}\, .
\end{equation}

\noindent That is, by the oscillatory term in the probability that the initial $\arrowvert K^0 \rangle $ is detected in the strangeness eigenstate $\arrowvert \overline{K}^{0} \rangle $ at the later time $\tau =0 $. Direct calculation gives

\begin{equation}\label{eq:vis_strange}
\mathcal{V}_0 (\tau ) = \dfrac{2e^{-\frac{1}{2}(\Gamma _S + \Gamma _L )\tau}}{e^{-\Gamma _S \tau } + e^{-\Gamma _S \tau } } \;\; .
\end{equation}
\vspace{0.1cm}
\par Next, a straightforward argument (see the Appendix) shows that the ratio $\frac{d\mathcal{V}_0}{d\tau }\Big/\frac{d\mathcal{V}}{d\tau }$ between its derivative and that of the fidelity visibility (\ref{eq:vis_fidelity}) is positive in the time interval $0\leq \tau \leq \tau _0 \,$. The quantities $\mathcal{V},\mathcal{V}_0 $ are then either both increasing or both decreasing in $[0,\tau _0 ]$. Therefore, we see from Eq.~(\ref{eq:inf_inequality}) that the increase of lifetime information as measured by $\mathcal{D}(\tau ) $ in fact enforces (not only $\mathcal{V} $, but also) the visibility $\mathcal{V}_0 $ of the strangeness oscillations to decrease in this interval.
%\vspace{-0.3cm}
%%%%%%%%%%%%%%%%%%%%%%%%%%%%%%%%%%%%%%%%%%%%%%%%%%%%%%%%%%%%%%%%%%%%%%%%%%%%%%%%%%%%%%%%%%%%%%%%%%%%%%%%%%%%%%%%%%%%%%%%%%%%%%%%%%%%%%%%%%%%%%
\section{Conclusions}
\label{sec:conclusions}

\par Entanglement plays a crucial role in quantum mechanical complementarity for bipartite systems.~We have shown in the present work how it can be clearly illustrated and discussed in the kaon-antikaon oscillating system.~We considered a bipartite model where a single neutral kaon interacts with the environment consisting of its weak interaction decay products.~From an interferometric point of view, the kaon is treated as the interfering object (quanton) and lifetime/width information plays the role of which-way information.~This is similar to the \emph{neutral kaon interferometry} of Bramon, Garbarino and Hiesmayr \cite{BGH_EPJC}. We verified that, as entanglement correlations are established between these two parties, lifetime information leaks and becomes available in the environmental state.~Corresponding to the entanglement generation and acquisition of lifetime information, we saw how the visibility of which-way interference is reduced. The interplay between the single-particle properties visibility/distinguishability and entanglement was proved to be governed by a quantitative complementarity relation:
\vspace{-0.3cm}
\begin{equation*}
\mathcal {V}^2 + \mathcal {D}^2 + \mathcal {S}^2 \leq 1
\end{equation*}

\noindent This inequality is similar to the one proposed by Jakob and Bergou in their analysis of wave-particle duality in bipartite systems \cite{Jakob_Bergou}.

%%%%%%%%%%%%%%%%% Fig. 3
\begin{figure}[htp]
	\begin{center}
		\includegraphics[scale=0.38]{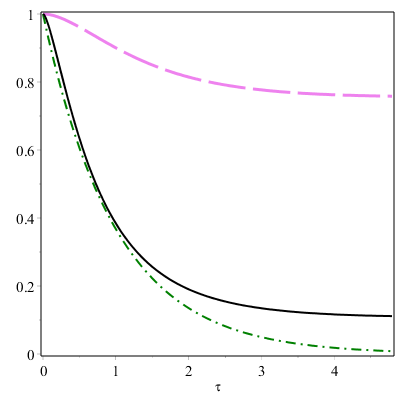}
		\caption{(colors online) Upper bounds for $\mathcal{V}^2 $ (the dotdashed green line) given by $1-\mathcal{D}^2 $ (the dashed purple line) and by $1-\mathcal {D}^2 -\mathcal {S}^2 $ (the solid black line).}\label{fig3}
	\end{center}
\end{figure}
%%%%%%%%%%%%%%%%%%
\vspace{-0.1cm}
\par In this direction, it is interesting to notice that the inclusion of the quantitative entanglement measure in this ``triality'' relation is very important if we want to see the reduction of the interference visibility $\mathcal{V}$ as enforced by quantum complementarity.~In Fig.~3, we compare the upper bounds for $\mathcal{V}^2 $ given by $1-\mathcal{D}^2 $ alone and by $1-\mathcal {D}^2 -\mathcal {S}^2 $. The upper bound including entanglement is much sharper and consistent with the actual reduction in $\mathcal{V}$.

\par We have also shown how our inequality accounts for the complementarity between strangeness oscillations and lifetime information in the time interval relevant for the analysis. This demonstrates consistency with the previous analysis of complementarity in the neutral kaon system, and with the general principle that the visibility of any quantum interference phenomenon whatsoever must reduce when which-way information becomes available \cite{Scully_Englert_Walther, Rempe,Englert_Scully_and_others}.
%\vspace{-0.3cm}
%%%%%%%%%%%%%%%%%%%%%%%%%%%%%%%%%%%%%%%%%%%%%%%%%%%%%%%%%%%%%%%%%%%%%%%%%%%%%%%%%%%%%%%%%%%%%%%%%%%%%%%%%%%%%%%%%%%%%%%%%%%%%%%%%%%%%%%%%%%%%%

%\section*{Acknowledgments}
%\vspace{1.0cm}
\textbf{Acknowledgments.} G.S. and M.S. would like to thank the Departamento de Ci\^{e}ncias Exatas e Tecnol\'{o}gicas/UESC -- Ilh\'{e}us for the hospitality and financial support during the development of this work. M.S. also acknowledges financial support form the Brazilian institutions CNPq (Conselho Nacional de Desenvolvimento Científico e Tecnológico) and FAPEMIG (Funda\c{c}\~{a}o de Amparo \`{a} Pesquisa do Estado de Minas Gerais). J.G.O. acknowledges financial support from FAPESB (Funda\c{c}\~{a}o de Amparo \`{a} Pesquisa do Estado da Bahia), AUXPE-FAPESB-3336/2014 number 23038.007210/2014-19.

%%%%%%%%%%%%%%%%%%%%%%%%%%%%%%%%%%%%%%%%%%%%%%%%%%%%%%%%%%%%%%%%%%%%%%%%%%%%%%%%%%%%%%%%%%%%%%%%%%%%%%%%%%%%%%%%%%%%%%%%%%%%%%%%%%%%%%%%%%%%%%
\section*{Appendix}

\par Let us show that the ratio $\frac{d\mathcal{V}_0}{d\tau }\Big/\frac{d\mathcal{V}}{d\tau }$ between the derivatives of the visibility of strangeness oscillations (Eq. (\ref{eq:vis_strange})) and the fidelity visibility (Eq. (\ref{eq:vis_fidelity})) is positive in the time interval $0\leq \tau \leq \tau _0 \,$.~Observe that this dimensionless ratio is given by

\begin{widetext}
\begin{equation*}
\frac{d\mathcal{V}_0}{d\tau }\Big/\frac{d\mathcal{V}}{d\tau } = \frac{2}{e^{-\Gamma _S \tau }+e^{-\Gamma _L \tau }}\left( 1 - \frac{\Gamma _S e^{-\Gamma _S \tau } + \Gamma _L e^{-\Gamma _L \tau }}{\Gamma (e^{-\Gamma _S \tau }+e^{-\Gamma _L \tau })} \right) > 0 \quad \forall 0\leq \tau \leq \tau _0 \, .
\end{equation*}
\end{widetext}

\noindent Therefore, it is enough to show that

\vspace{-0.2cm}
\begin{equation*}\label{eq:ratio}
\frac{\Gamma _S e^{-\Gamma _S \tau }+\Gamma _L e^{-\Gamma _L \tau }}{\Gamma \left( e^{-\Gamma _S \tau }+e^{-\Gamma _L \tau } \right) } \leq 1\, , \quad \forall \, 0\leq \tau \leq \tau _0 \, .
\end{equation*}

\par In order to do this, notice that since $\Gamma _S = 579\Gamma _L $, we have $e^{-\Gamma _S \tau }\leq e^{-\Gamma _L \tau }$ for every $0\leq \tau \leq \tau _0 $. Thus, in the interval $[0,\tau _0 ]$ we have $$(\Gamma _S -\Gamma _L )e^{-\Gamma _S \tau }\leq (\Gamma _S -\Gamma _L )e^{-\Gamma _L \tau },$$ \noindent or

\begin{equation*}
\frac{1}{2}\left( \Gamma _S e^{-\Gamma _S \tau }+\Gamma _L e^{-\Gamma _L \tau } \right) \leq \frac{1}{2}\left( \Gamma _L e^{-\Gamma _S \tau }+\Gamma _S e^{-\Gamma _L \tau } \right)\; .
\end{equation*}

\noindent Adding $\frac{1}{2}\left( \Gamma _S e^{-\Gamma _S \tau }+\Gamma _L e^{-\Gamma _L \tau } \right)$ to both sides of the previous inequality gives

\begin{equation*}
\Gamma _S e^{-\Gamma _S \tau }+\Gamma _L e^{-\Gamma _L \tau }\leq \Gamma \left( e^{-\Gamma _S \tau }+ e^{-\Gamma _L \tau } \right)\; ,
\end{equation*}

\vspace{0.2cm}
\noindent as desired.
\newpage
%%%%%%%%%%%%%%%%%%%%%%%%%%%%%%%%%%%%%%%%%%%%%%%%%%%%%%%%%%%%%%%%%%%%%%%%%%%%%%%%%%%%%%%%%%%%%%%%%%%%%%%%%%%%%%%%%%%%%%%%%%%%%%%%%%%%%%%%%%%%%%


\begin{thebibliography}{99}
\bibitem{footnote_1} The lifetimes $\tau _S $, $\tau _L $ of the $K_S^0 $, $K_L^0 $ neutral kaons are $\tau _S = 8.958\times 10^{-11}\mathrm{seg}$, $\tau _L \approx 579 \tau _S $. The $K_{S}^{0}$--$K_{L}^{0}$ mass gap is $\Delta m = m _L - m_S = 3.5 \times 10^{-12}\,\mathrm{MeV} = 5.3 \times 10^{9}\,\mathrm{seg}^{-1}$.
\bibitem{footnote_3} The Wigner-Weisskopf prescription is to take $\arrowvert K_{S}^{0}(\tau )\rangle = e^{-im_S \tau }e^{-\frac{1}{2}\Gamma _S \tau }\arrowvert K_{S}^{0} \rangle $ and $\arrowvert K_{L}^{0}(\tau )\rangle = e^{-im_L \tau }e^{-\frac{1}{2}\Gamma _L \tau }\arrowvert K_{L}^{0} \rangle $.
\end{thebibliography}
\end{document}